%% file: PARP.tex
\documentclass[conference]{IEEEtran}
\IEEEoverridecommandlockouts

\usepackage{graphicx}
\usepackage{textcomp}
\usepackage{xcolor}
\def\BibTeX{{\rm B\kern-.05em{\sc i\kern-.025em b}\kern-.08em
    T\kern-.1667em\lower.7ex\hbox{E}\kern-.125emX}}

\usepackage{cite}
\usepackage{amsmath,amssymb,amsfonts}

\usepackage{graphicx}
\graphicspath{ {./images/} }
\usepackage{textcomp}
\usepackage{xcolor}

\usepackage{tabularx,booktabs}
\newcolumntype{C}{>{\centering\arraybackslash}X} 
\usepackage{lipsum} 

\usepackage{caption}

\usepackage{hyperref}

\usepackage{enumitem} 

\usepackage{algorithm} 
\usepackage{algpseudocode}

\usepackage{array}
\usepackage{booktabs}

\usepackage{graphicx}
\usepackage{threeparttable}
\usepackage{wasysym}

\newcommand{\res}{\mathit{res}}
\newcommand{\req}{\mathit{req}}

\newcommand{\PC}{\mathcal{P}}
\newcommand{\channelId}{\alpha}
\newcommand{\FN}{\mathit{FN}}
\newcommand{\pkfn}{\mathit{pk}_{\FN}}
\newcommand{\skfn}{\mathit{sk}_{\FN}}
\newcommand{\LC}{\mathit{LC}}
\newcommand{\WN}{\mathit{WN}}
\newcommand{\pklc}{\mathit{pk}_{\LC}}
\newcommand{\sklc}{\mathit{sk}_{\LC}}

\newcommand{\budget}{\mathit{b}}
\newcommand{\closingState}{\mathit{cs}}
\newcommand{\status}{\mathit{T}}

\newcommand{\blockHash}{\mathit{h_B}}
\newcommand{\paymentAmount}{\mathit{a}}
\newcommand{\rpcCall}{\gamma}
\newcommand{\hashReq}{h_{\req}}
\newcommand{\hashRes}{h_{\res}}
\newcommand{\sigReq}{\sigma_{\req}}
\newcommand{\sigAmount}{\sigma_{a}}

\newcommand{\blockHeight}{\mathit{m_B}}
\newcommand{\rpcResult}{\mathit{R(\gamma)}}
\newcommand{\rpcProof}{\mathit{\pi_\gamma}}
\newcommand{\sigRes}{\sigma_{\res}}

\begin{document}

\title{Depermissioning Web3: a Permissionless Accountable RPC Protocol for Blockchain Networks}


\author{\IEEEauthorblockN{Weihong Wang}
\IEEEauthorblockA{\textit{DistriNet, KU Leuven} \\
Leuven, Belgium \\
weihong.wang@kuleuven.be}
\and
\IEEEauthorblockN{Tom Van Cutsem}
\IEEEauthorblockA{\textit{DistriNet, KU Leuven} \\
Leuven, Belgium \\
tom.vancutsem@kuleuven.be}}

\maketitle
\thispagestyle{plain}
\pagestyle{plain}

\begin{abstract}

\input{Content/0-Abstract}

\end{abstract}







\input{Content/1-Introduction}

\input{Content/3-S_P-issues}

\input{Content/2-Background}

\input{Content/4-Setup-and-Goals}

\input{Content/5-Main-Protocol}
\input{Content/6-Implementation}

\input{Content/7-Related-work}

\input{Content/8-Future-work}

\input{Content/9-Conclusion}

\section*{Acknowledgments}
We thank the anonymous reviewers for their constructive feedback, which  improved the clarity and quality of this work. We are also grateful to Glenn, Jan, and Kristof for their helpful comments and suggestions on the manuscript. This research was partially funded by the Research Fund KU Leuven and the Cybersecurity Research Program Flanders. 

\bibliographystyle{IEEEtran}
\bibliography{biblio}

\end{document}

%% file: Content/0-Abstract.tex
In blockchain networks, so-called ``full nodes'' serve data to and relay transactions from clients through an RPC interface. This serving layer enables integration of ``Web3'' data, stored on blockchains, with ``Web2'' mobile or web applications that cannot directly participate as peers in a blockchain network. In practice, the serving layer is dominated by a small number of centralized services (``node providers'') that offer permissioned access to RPC endpoints. Clients register with these providers because they offer reliable and convenient access to blockchain data: operating a full node themselves requires significant computational and storage resources, and public (permissionless) RPC nodes lack financial incentives to serve large numbers of clients with consistent performance.

Permissioned access to an otherwise permissionless block\-chain network raises concerns regarding the privacy, integrity, and availability of data access. To address this, we propose a \textbf{P}ermissionless \textbf{A}ccountable \textbf{R}PC \textbf{P}rotocol (PARP). It enables clients and full nodes to interact pseudonymously while keeping both parties accountable. PARP leverages ``light client'' schemes for essential data integrity checks, combined with fraud proofs, to keep full nodes honest and accountable. It integrates payment channels to facilitate micro-payments, holding clients accountable for the resources they consume and providing an economic incentive for full nodes to serve. Our prototype implementation for Ethereum demonstrates the feasibility of PARP, and we quantify its overhead compared to the base RPC protocol.





%% file: Content/1-Introduction.tex
\section{Introduction}\label{section:introduction}
Blockchain networks support decentralized applications or ``dApps'' by storing data on a peer-to-peer (P2P) network, as opposed to traditional ``Web2'' applications supported by centralized server-centric data storage.
This shift introduces complexities for application clients because participating directly in a blockchain network's P2P protocol is challenging. Not only does it require the client to act as a server, but it also requires the client to potentially maintain a full copy of the network state. However, the computational and storage demands of running a so-called ``full node'' are considerable (e.g., Ethereum nodes require 2TB of SSD storage and 25 MBit/s bandwidth~\cite{ethreq}). Clearly, it is not practical for all end-users to set up such infrastructure.

Additionally, due to the resource-intensiveness of operating a full node, combined with the lack of any economic incentives, the owners of those full nodes are hesitant to provide end-users with free and unrestricted access to blockchain data through their nodes' RPC interfaces, especially as the demand and number of application clients grows larger.

\begin{figure}[htbp] 
    \centering 
    \vspace{-5pt} 
    \includegraphics[width=0.8\linewidth]{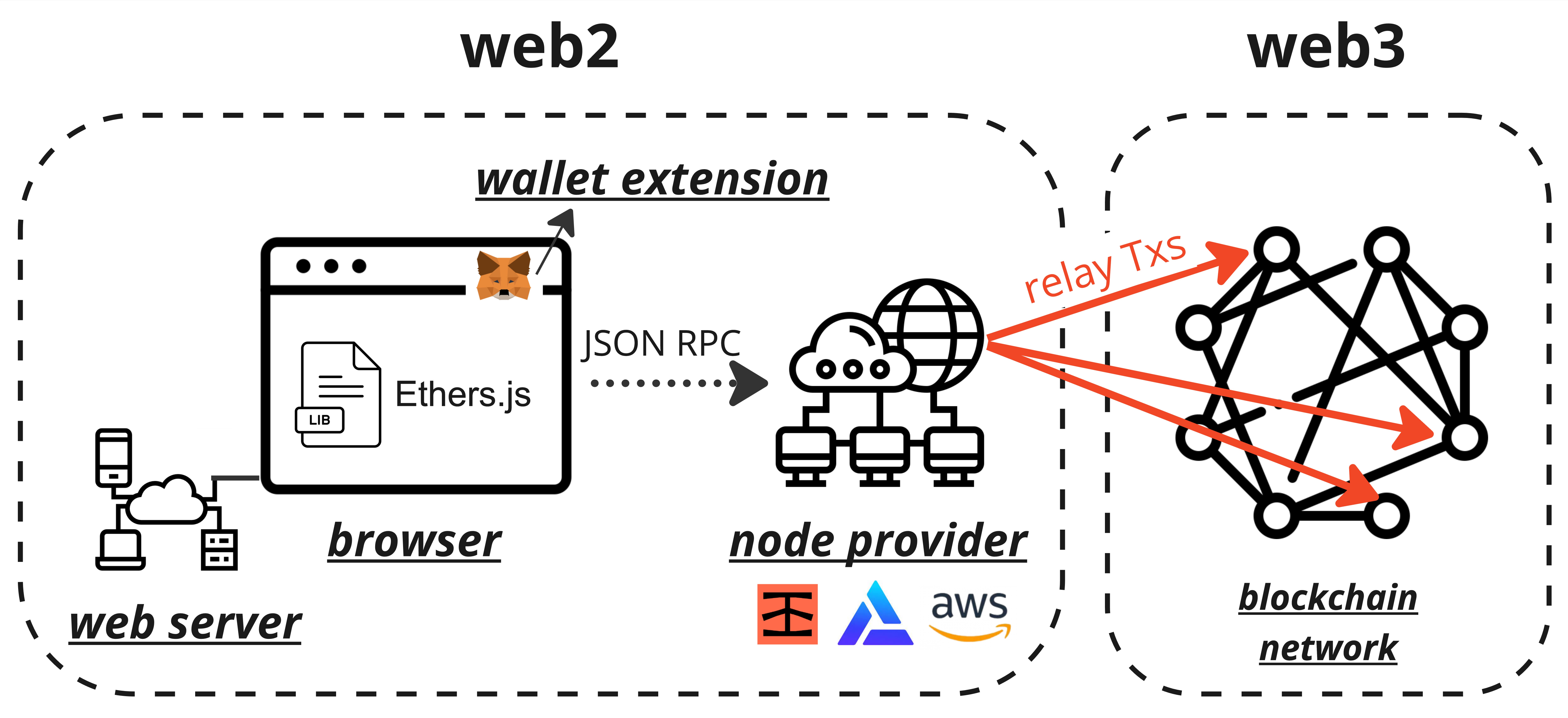}
    \caption{A schematic of the typical serving layer underlying decentralized applications (dApps) using Web3 protocols.}
    \label{fig:web3-ps}
    \vspace{-3pt} 
\end{figure}

Consequently, many resource-limited clients rent access to full nodes through node-as-a-service (NaaS)~\cite{ethereumnaas} providers or simply ``node providers''. Popular node providers include Infura~\cite{infura} and Alchemy~\cite{AlchemyWeb}, which operate full nodes on several blockchain networks and offer hosted RPC services. The NaaS trend emerged as a convenient solution for users to interact with Web3 protocols. Figure~\ref{fig:web3-ps} illustrates a typical scenario where a dApp connects to a node provider, which relays user transactions to the blockchain network and retrieves data from the blockchain for presentation to the user. 

The connection between applications and nodes run by node providers involves an API key, which authenticates who the user of this key is. By tracking and controlling how the API is used, node providers can charge the user based on their pricing plans. In the majority of cases, applying for an API key requires users to undergo a registration process, which may include sharing sensitive information such as personal identification or financial information.


The reliance on a trusted intermediary to interact with a trustless, decentralized network runs counter to the core principles of why Web3 protocols were invented in the first place~\cite{MM22}. In particular, users often trade \textbf{privacy}, \textbf{integrity}, and \textbf{availability} for convenience, much like they did in ``Web2''. Specifically: 
\begin{itemize}
    \item \textbf{Privacy.} When registering at a NaaS provider, users have to authorize the collection of personally identifiable information (PII), sometimes including their full name and credit card details, making it easy to profile users.

    \item \textbf{Integrity.} Node providers can manipulate or misrepresent blockchain data without penalties~\cite{MM22}, leaving users without hard guarantees that the retrieved data is in accord with the latest data stored on the blockchain. 

    \item \textbf{Availability.} Node providers face regulatory implications from laws in their operating countries, potentially leading to censorship of sanctioned applications and users in specific regions~\cite{InfuraReg}.  
    
\end{itemize}




The blockchain RPC serving layer lacks accountability and economic incentives to provide applications and end-users with reliable access to on-chain data. In Ethereum, for example, full nodes receive financial rewards for validating transactions using ``Proof-of-Stake'' consensus and by charging fees on validated transactions, but not for serving RPC requests or relaying transactions. Misbehavior as a validator (e.g., by attesting to an incorrect block in Ethereum) is discouraged through slashing conditions, but no such mechanism exists for misbehavior in the serving layer. Yet, both validation and serving play a crucial role in a healthy network.

In short, Web3 protocols heavily rely on economic incentives to keep nodes honest and participatory, while such incentives are missing for the serving layer. We aim to address this gap by introducing mechanisms in the serving layer that enable (proportional) financial rewards for nodes that serve and, at the same time, ensure detection and punishment of misbehavior towards clients.



To introduce these mechanisms, we designed PARP, a \textbf{P}ermissionless \textbf{A}ccountable \textbf{R}PC \textbf{P}rotocol that wraps a block\-chain network's base layer RPC protocol with additional layers of authentication and accounting. PARP aims to enhance the end users' access to reliable blockchain data while at the same time compensating full nodes fairly for the service provided. PARP leverages known light client schemes to improve data integrity for the clients. In addition, it introduces payment channels between clients and full nodes to allow full nodes to accept micro-payments from clients while safeguarding the pseudonymity of both parties. To act as a PARP server and accept such micro-payments, a full node must lock up funds that can be slashed on misbehavior, thus deterring malicious actors in the serving layer.





Our main contributions are the following:
\begin{itemize}
    \item We propose a new serving layer model that complements the current endpoint infrastructure provided by permissioned services (NaaS providers). It features accountability mechanisms and incentives that mutually benefit clients and full nodes.
    \item We propose a design that safeguards the end users data security and does not require trust in a centralized third party.
    \item We implement PARP for the Ethereum RPC layer and integrate it into a widely used Ethereum execution node implementation (Geth), providing evidence for the feasibility of the protocol and its compatibility with current blockchain networks. 
\end{itemize}

%% file: Content/3-S_P-issues.tex
\section{Web3 Serving Layer Issues}

\input{table/comparison.tex}

In this section, we discuss key considerations in the Web3 serving layer, including insufficient economic incentives for full nodes, dominance of major node providers and how their permissioned registration processes enable extensive profiling, and the trade-offs between accountability and permissionlessness in accessing blockchain networks.

\subsection{Lack of Economic Incentives for Full Nodes}

As light clients emerge to address heavy resource requirements for devices like smartphones, they face the challenge of bootstrapping and synchronizing themselves securely and efficiently~\cite{CBC22}. However, serving requests from many light clients places a substantial burden on these full nodes regarding network resources and communication. Therefore, with an increasing number of light clients benefiting from enhanced privacy and security joining the network, the establishment of effective incentives to compensate full nodes for supporting light clients becomes crucial.

\subsection{Centralization among Node Providers}
To evaluate the role of node providers, we analyzed a dataset \cite{torresdataset} from Torres \textit{et al.} \cite{Torres23walletPrivacy}, which contains detailed records of web traffic from 1572 dApps. From this, we specifically focused on 383 dApps that send JSON-RPC calls directly from their frontend to node providers for blockchain data access. We then mapped these JSON-RPC calls to identify which node providers each dApp interacts with, noting that a single dApp can connect to multiple providers. This allows us to determine the extent of traffic received by each provider. The results indicate that \textbf{47.52\%} of the dApps in our dataset connect to Infura~\cite{infura}, making it the most widely used provider by a significant margin. Alchemy~\cite{AlchemyWeb} is the second-most common, used by \textbf{31.07\%} of dApps, followed by Binance (12.01\%), Ankr (9.4\%), Cloudflare (6.79\%) and other providers. 

\subsection{Permissioned Nature of Node Providers}

We picked five top node providers from our dataset, excluding network-specific ones. We examined their traits on their websites during the API key application process, as shown in Table \ref{tab:features}.

First, we inspected the \textbf{registration requirements} of each provider to see if their services could be used without registration. Ankr stands out as the only provider that has a list of free endpoints.

For \textbf{wallet-based identity}, only Ankr supports this, granting users flexibility in service access. Conversely, the remaining providers ask for at least an email address for registration.

We also assessed \textbf{free monthly requests} and the pricing models of each provider. All providers offer some free daily service to users, while 3 out of 5 charge based on varied call types for a fairer fee calculation.

Finally, we reviewed the accepted \textbf{payment methods}, finding that 2 out of 5 providers accept crypto.

In conclusion, end users still need to register for an account to access a reliable blockchain connection, much like in Web2 services. In combination with all the Web3 requests sent to these node providers, it becomes fairly easy for them to construct user profiles.

\subsection{Tradeoffs Between Accountability and Permissionlessness}

Today clients can access some public RPC endpoints, either from node providers or anonymous full nodes~\cite{awesome-np, chainlist}, but each option comes with its own tradeoffs. For node providers, access is permissioned due to registration requirements, but users can put some trust in the data as node providers have a reputation and commercial interests to uphold. By contrast, public anonymous RPC endpoints are permissionless, but there is no accountability for these anonymous nodes to serve information reliably and correctly. Hence, achieving both permissionlessness and accountability in the serving layer remains an open issue.

%% file: table/comparison.tex

\begin{table*}[t]
\footnotesize
\newcommand*\rot[1]{\hbox to1em{\hss\rotatebox[origin=br]{-60}{#1}}}
\newcommand*\feature[1]{\ifcase#1 -\or\LEFTcircle\or\CIRCLE\fi}
\newcommand*\f[3]{\feature#1&\feature#2&\feature#3}
\newcommand*\ff[3]{&\feature#2&\feature#3}
\newcommand*\ft[3]{&\feature#2&#3}

\makeatletter

\newcommand*\ex[8]{#1 & \feature#2 & {\kern1em} \feature#3 {\kern2em} &  \f#4 &  {\kern2em} \ft#5 & #6 & \ff#7 & #8 \expandafter\@firstofone}
\makeatother
\newcolumntype{G}{c@{}c@{}c}
\newcolumntype{F}{c@{}c}

\begin{threeparttable}
\caption{Comparison of Features and Registration Requirements of Node Providers\textsuperscript{‡}}
\label{tab:features}
\begin{tabularx}{\textwidth}{ c  c c  G  G  r G c c}
\toprule

Node Provider  & \multicolumn{1}{c}{Free Public Service} & \multicolumn{1}{c}{Login}  & \multicolumn{3}{c}{Sign-up} &  \multicolumn{3}{c}{Pricing Plan} & Freemium Node Service & \multicolumn{3}{c}{Payment} & \multicolumn{1}{c}{Traffic Share}
\\
\midrule

 & \rot{No Signup}
 
 & \rot{Via wallets}
 
 & \rot{Email}
 & \rot{Full name}
 & \rot{Org name}

  & \rot{}
 & \rot{Call-Based}
 & \rot{Plan tiers}

 & \rot{Free usage\textsuperscript{†}}
 
   & \rot{}
 & \rot{Credit card}
 & \rot{Crypto}
 \\
\midrule
\ex{Infura~\cite{infura}}      {0}  {0} {220}     {-05} {3 million credits (daily)}            {-20}   {\makebox[3em][r]{182/383} (47.52\%)} \\
\ex{Alchemy~\cite{AlchemyWeb}}      {0}  {0} {220}     {-24} {300 million compute units (monthly)}     {-20}  {\makebox[3em][r]{119/383} (31.07\%)} \\

\ex{Ankr~\cite{ankr}}         {2}  {2\textsuperscript{*}} {100}     {-04} {30 requests (per sec)}           {-22} {\makebox[3em][r]{36/383} (9.4\%)}\\

\ex{Quicknode~\cite{quicknode}}    {0}  {0} {222}     {-25} {10 million API credits (monthly)}        {-20} {\makebox[3em][r]{16/383} (4.18\%)}\\

\ex{Chainstack~\cite{chainstack}}   {0}  {0} {222}     {-24} {3 million request units (monthly)}       {-22}  {\makebox[3em][r]{5/383} (1.31\%)} \\


\bottomrule
\end{tabularx}
\begin{tablenotes}
\item \hfil$\feature2=\text{indicates that the property is provided; in the Sign-up column, it means the property is required.}$; 
\item \hfil$\feature1=\text{indicates that property is not necessarily required}$; \hspace{0.5em} $\text{\feature0}=\text{does not provide property}$. 
\item[*] Wallets must have active transactions in the past to be supported by the node provider. \hspace{5em} \textsuperscript{‡} All the data was collected before December 2024. 
\item[†] These metrics (compute units, requests, credits, etc.) represent the amount of computational work and resource usage as defined by different node providers. 
\end{tablenotes}    

\end{threeparttable}

\end{table*}


%% file: Content/2-Background.tex
\section{Background}
In this section, we discuss the state of practice in bridging Web2 and Web3 protocols. We also provide the necessary background on light client schemes and payment channels.

\subsection{Endpoint Infrastructure}

To interact with the blockchain network, clients connect to a full node via universally supported APIs like JSON-RPC.\ An RPC endpoint is the network location where an application sends its API requests to a full node for execution. The endpoint typically provides access to various functionalities of the blockchain network, including broadcasting transactions and retrieving blocks and block headers.

Most dApp developers obtain an API key with a node provider for the network they wish to use (as discussed in the Introduction~\ref{section:introduction}). DApp end-users often can't configure the underlying API keys that their dApps are using. Although many wallets do allow end-users to configure their own RPC URL, most opt to use the default settings, commonly referring to a node provider url. For example, MetaMask, a widely-used Web3 wallet for Ethereum uses Infura~\cite{MMWI} as its default endpoint provider to query the Ethereum blockchain and obtain the balance for the end-user's addresses.


\subsection{Light Client Solutions}

Light client schemes, such as simple payment verification (SPV)~\cite{SN08}, offer a compromise between the resource-intensive demands of operating a full node and the security risks of relying on third-party servers.

\textbf{Efficiency.} Light clients download only block headers, significantly reducing data requirements (e.g., an 80-byte header vs.\ a 1MB full block). Schemes, such as FlyClient~\cite{FlyClient} and Coda~\cite{bonneau2020coda}, achieve fast block header synchronization even up to a constant size to the length of the chain. 


\textbf{Data integrity.} By leveraging Merkle trees, light clients can verify transactions or information against the transaction root or state root of the tree contained in the block header.

\textbf{Reliance on full nodes.} Due to storage limitations of light clients, they still rely on their peer full nodes to obtain essential information to keep up-to-date with the tip of the chain. 

\subsection{Payment channels}


Payment channels were proposed to enable micropayments between two parties without recording each individual payment as a transaction on the blockchain~\cite{Bitcoin1PayChan}, thus \textbf{increasing throughput} and \textbf{reducing costs} (fees).


A payment channel operates under predefined rules set by a smart contract~\cite{SokL2}. It is opened with one on-chain transaction in which both parties deposit funds, indicating the total amount available for ``off-chain'' transactions. The parties then manage a local off-chain ledger to track their balances. Upon closing the channel, if both agree on the final balance, funds are settled accordingly on the blockchain. In case of disputes, the smart contract acts as a mediator, verifying the provided evidence and disbursing funds based on the validated final state.



%% file: Content/4-Setup-and-Goals.tex
\section{PARP Protocol Overview}

We describe the design goals of PARP, introduce terminology and assumptions specific to our protocol, and then describe the full lifecycle of a PARP connection.

\subsection{Overview and Design Goals}\label{Sec4-overview}

PARP is an RPC protocol designed to allow a \emph{light client} to interact with a blockchain through a \emph{full node} in such a way that the interaction is both \emph{permissionless}, \emph{accountable} and \emph{economically sustainable} for both parties.

To participate as a full node in PARP, the node's operator must first deposit tokens as collateral to incentivize honest behaviour. While alternative approaches to accountability exist, PARP adopts a collateral-based mechanism to provide clear and quantifiable assurances for both parties. 

To receive service from a PARP full node, a PARP light client must first open a payment channel and commit to deposit funds in the channel before it can start making RPC requests. These funds represent the light client's budget to pay the full node for its service. Additionally, the PARP protocol allows the light client to detect incorrect RPC responses and report them (via another full node) through a fraud-proof protocol, penalizing the misbehaving PARP full node (by withholding part of its collateral).

The design goals of PARP include:

\begin{enumerate}
    \item Accountability:
        \begin{itemize}
            \item \textbf{Bilateral assurances.} Deposits by light clients and full nodes establish mutual accountability, ensuring reliable services and fair compensation within the protocol.
            \item \textbf{Trust and verification.} By leveraging merkle proofs, light clients can verify the correctness of the data returned in an RPC response. Together with a fraud-proof protocol, the full node is incentivized to serve RPC requests correctly.
        \end{itemize}
    \item Permissionlessness:
        \begin{itemize}
            \item \textbf{Pseudonymity.} Our protocol reduces the leakage of personally identifying information by enabling end-users and dApps to interact with blockchains without requiring registration and without API keys that can correlate requests and build up user profiles. 
            \item \textbf{Enhanced availability.} PARP full nodes are discoverable via an on-chain registry and because of the lack of any sign-up process, clients can trivially switch between different PARP full nodes, e.g., for fail-over.
        \end{itemize}
    \item Economic incentives:
        \begin{itemize}
            \item \textbf{Cost-efficient payments.} Micropayments in a payment channel are used to compensate full nodes for their services, incentivizing them to serve light clients. The funds deposited by a light client are specifically allocated for payments to a full node within the channel.
        \end{itemize}
\end{enumerate}

\subsection{Roles}

The protocol involves three main roles:

\textbf{\textit{Full Node}}: A computer running the PARP-compatible full node software and connecting to other (PARP or non-PARP compatible) full node peers. It stores complete blockchain data, though in practice, it may be periodically pruned to save disk space~\cite{vb15pruned}. To serve as a full node, it must deposit tokens to a PARP-specific smart contract as collateral. 

\textbf{\textit{Light Client}}: A computer with limited storage, compute and network resources. A light client downloads only \textbf{block headers} rather than full blocks. It interacts with its connected PARP full node by making RPC requests, manages essential chain information, verifies data integrity using Merkle proofs against the root hash in block headers. It generates fraud-proofs if it detects invalid RPC responses.

\textbf{\textit{Payment Channel}}: A unidirectional ledger payment channel set up between a light client and a full node, signifying a successful connection between the two parties. A light client must deposit funds into this channel on-chain, representing the total amount of tokens it will pay a full node for its services. 

\subsection{On-Chain Modules}

The protocol also includes three additional on-chain modules deployed as smart contracts:

\textbf{\textit{Full Nodes Deposit Module}}: This module enables a full node to deposit its tokens, making it eligible to serve light clients in the network.

\textbf{\textit{Channels Management Module}}: This module manages payment channels for each PARP connection. It is responsible for managing the state of the payment channels, including the allocation of funds between the participating light client and the full node upon connection closure. Each payment channel has a unique identifier, based on the identity of the participants, which is recorded in this module.

\textbf{\textit{Fraud Detection Module}}: This module processes fraud proofs from light clients. If verified, it penalizes misbehaving full nodes by slashing their collateral and rewards the reporting participants, as detailed in Section~\ref{section4:fraudproof}.

\begin{figure}[htbp] 
    \centering 
    \vspace{-2pt} 
    \includegraphics[width=0.9\linewidth]{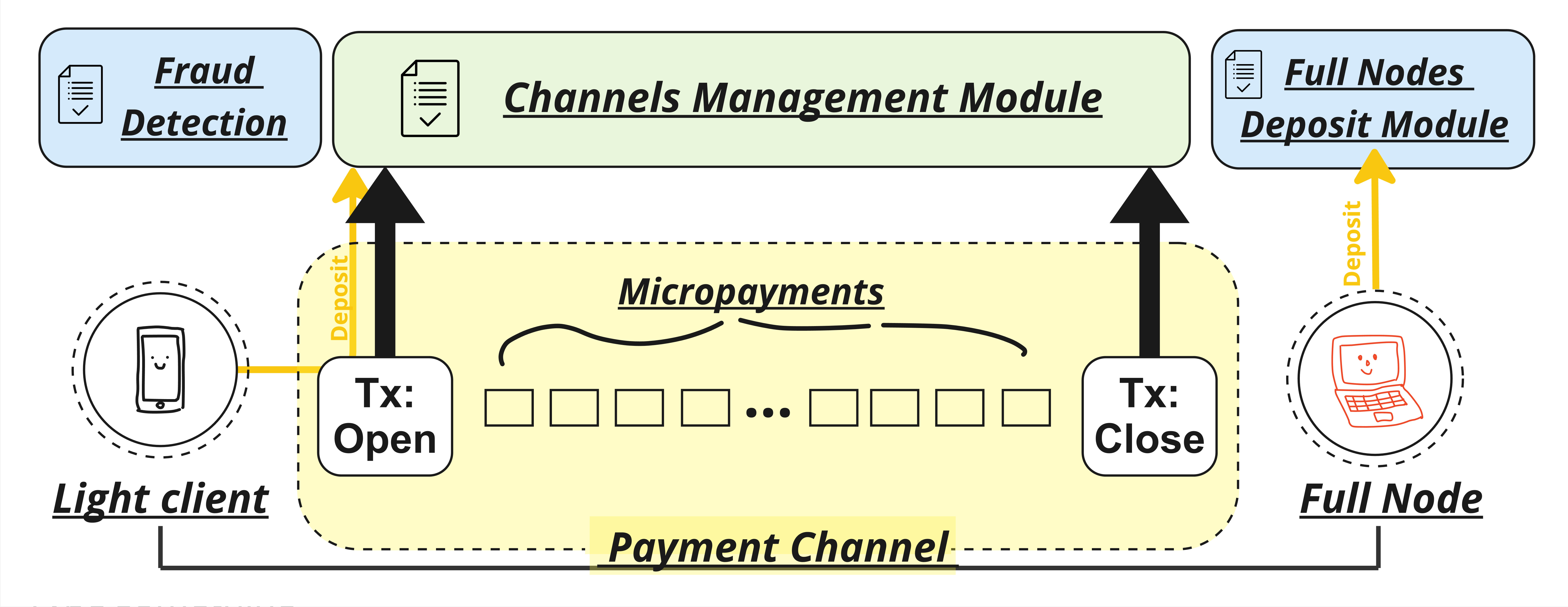}
    \caption{Primary participants in a PARP connection.}
    \label{fig:overview}
    \vspace{-5pt} 
\end{figure}

Figure~\ref{fig:overview} illustrates how all the participants in the PARP protocol interact.

\subsection{Assumptions}\label{subsec:assumption}

A light client relies on trusted sources to remain updated with the latest block headers in the blockchain~\cite{SN08, bonneau2020coda, FlyClient}. We assume that the light client can request and receive \textbf{block header}, which is compact, not client-specific, and serves as the root of trust, from \textit{any} full node (PARP-compatible or not), without payment. Access to block headers enables the light client to independently verify data integrity and network liveness. 

Furthermore, our protocol assumes a \textbf{strong synchrony} model. Specifically, we assume that messages between honest parties are delivered within a bounded delay. 

\subsection{Settings and Lifecycle of a PARP Connection}


\subsubsection{\textbf{Bootstrapping}}
Before initiating a PARP connection, the light client retrieves block headers from full nodes summarizing the blockchain state, including the chain tip and relevant Merkle roots. It then handshakes with a selected PARP-compatible full node to agree on connection parameters.

\subsubsection{\textbf{Connection Setup}}
The connection is created through an on-chain transaction to open a payment channel. This transaction is authorized by the light client and mediated via the full node, and includes the transfer of a client deposit, and the (pseudonymous) identity of both participants.

\subsubsection{\textbf{Request and Response phase}}
Once the payment channel is created, light client and full node can transact ``off-chain''.
Light clients can send any number of \textit{Request} messages, and full nodes respond with \textit{Response} messages. 

When a light client initiates a Request, it must include a micropayment, which indicates the payment amount, along with a blockchain-specific RPC call (e.g., for Ethereum, a call like \textit{eth\_getBalance(address)}), and other parameters. The payment amount is cumulative, meaning it adds up the total amount owed by the light client for all previous calls along with the current call.
Subsequently, the full node responds to the request, and both parties retain relevant records. Both the light client and the full node primarily track the requests, as each request contains a signed cumulative payment amount that enables the full node to redeem these funds. The response, recorded by the light client, may be used to generate a fraud-proof, as discussed in Section~\ref{section4:fraudproof}.

\subsubsection{\textbf{Closure (with or without dispute)}}

The connection closure can be triggered by either party through another on-chain transaction. The transaction must include the latest signed payment amount to close the channel. The channel will have a dispute window for a period of time before it closes.

\textit{Closure without dispute.} The final state of the channel is settled on-chain. The funds are distributed accordingly: the remaining unspent budget is returned to the light client, and the full node receives payment. Once this process is completed, the channel is settled, and the connection is closed. 

\textit{Closure with dispute.} Either party can present the latest state of the channel, represented by a payment proof (a signed message from the light client) with the largest cumulative payment amount. After the dispute period ends and the dispute is resolved based on the validity of the submitted payment proof, the channel is settled, and the connection is closed. 

\subsection{Fraud-Proof Protocol}
\label{section4:fraudproof}


The light client verifies the response from the full node by applying several checks. Based on the checks, the response is treated as: 

\begin{itemize}
    \item \textbf{Valid}: All checks pass successfully; the client trusts it.
    \item \textbf{Invalid}: The client cannot trust the response, but also cannot hold the full node accountable for fraud. It is sensible for the client to terminate the connection.
    \item \textbf{Fraudulent}: The client cannot trust the response, terminates the connection, and can take steps to construct a fraud proof to penalize the full node for its misbehavior.
\end{itemize}

In the case of fraud, to penalize the full node, the client must submit a fraud proof to the \textit{Fraud Detection Module} contract. Obviously we cannot trust the full node to submit a proof of its own fraudulent behavior to the blockchain. The light client must instead resort to another PARP-compatible full node, which we refer to as a \textbf{witness full node}. 



To verify the fraud proof, the \textit{Fraud Detection} contract can use the request and response data to re-check all the conditions stated above. 





If the fraud proof is deemed valid by the \textit{Fraud Detection Module}, the contract will instruct the \textit{Deposit Module} to confiscate the deposit of the full node and distribute it to three parties: the network's serving layer nodes (to incentivize punishment of fraudulent nodes), the light client (to incentivize reporting of fraudulent nodes), and the witness full node (to incentivize assistance in reporting fraudulent nodes).
The witness node is compensated by the contract directly. The light client does not need to establish a payment channel with the witness node.

%% file: Content/5-Main-Protocol.tex
\section{PARP Design Details}
\label{chpt:main-protocol}

This section outlines the network assumptions, introduces the protocol states and messages, and further illustrates the state transition within participants in a PARP connection. It also discusses the liveness check on payment channels and fraud detection mechanisms.

\subsection{Protocol States and Messages}~\label{subsec:protocolStateAndMessages}

A full node is denoted as $\FN$, a light client as $\LC$, and a payment channel as $\PC$.  Let $(\pkfn, \skfn)$ and $(\pklc, \sklc)$ denote their public/private key pairs, respectively, with $\mathit{addr}$ denoting their corresponding addresses.





\begin{figure}[htbp] 
    \centering 
    \includegraphics[width=\linewidth]{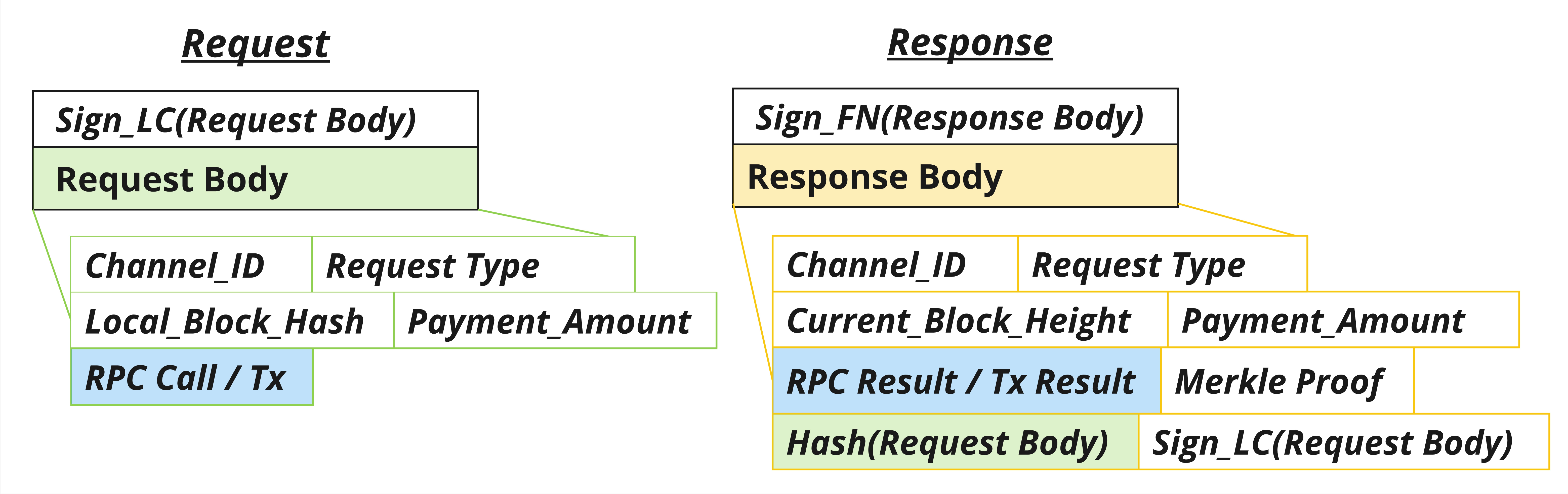}
    \caption{Structure of a PARP request and a PARP response}
    \label{fig:reqres}
\end{figure}

Within the off-chain connection, a light client's request to a full node is denoted as $\req$, while a corresponding response is represented as $\res$. Both the request and response in each RPC call round $i$ must be signed by their respective parties before being sent. The data structure is defined in Figure~\ref{fig:reqres}.

\begin{figure*}[t] 
    \centering 
    \includegraphics[width=0.9\textwidth]{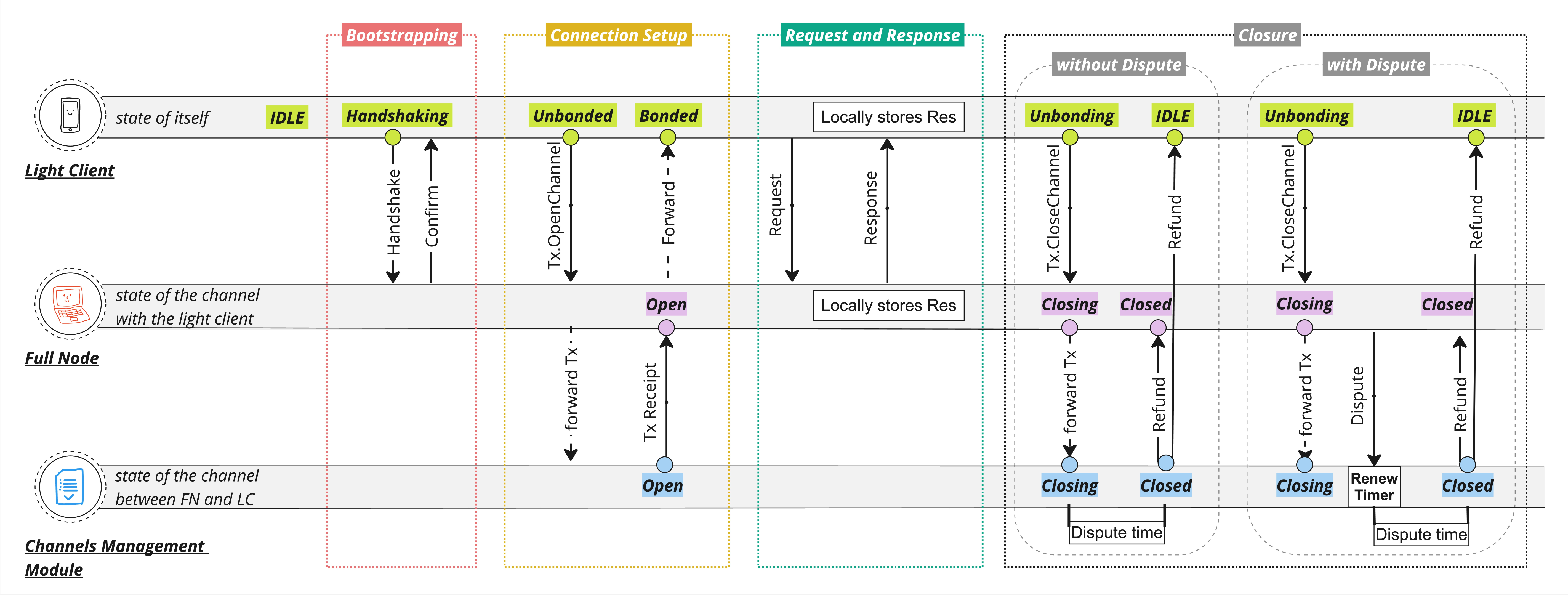}
    \caption{Tripartite state transition diagram illustrating the lifecycle of a PARP connection.}
    \label{fig:state}
\end{figure*}

Let the request be denoted by \( \req \) and defined as follows:
\[
\req = (\channelId, \blockHash, \paymentAmount, \rpcCall, \hashReq, \sigAmount, \sigReq)
\]
where:
\begin{itemize}
    \item $\channelId$ is the identifier of the channel between $\LC$ and $\FN$.
    \item $\blockHash$ is the most recent valid hash of a blockchain block, denoted as $B$, stored by $\LC$.
    \item $\paymentAmount$ indicates the amount $\LC$ is willing to pay for all previous calls along with the current call. It should satisfy $req_i.\paymentAmount \geqslant req_{i-1}.\paymentAmount$, indicating cumulative payments.
    \item $\rpcCall$ is an RPC call that $\LC$ wants to get executed.
    \item $\hashReq$ is the result of $\mathit{Hash}(\channelId, \blockHash, \paymentAmount, \rpcCall)$.
    \item $\sigAmount$ is $\mathit{Sign}(\mathit{Hash}(\channelId, \paymentAmount), \sklc)$, for payments. 
    \item $\sigReq$ is $\mathit{Sign}(\hashReq, \sklc)$, for verification.
\end{itemize}





And the response is denoted as $\res$:
\[
\res = (\channelId, \blockHeight, \paymentAmount, \rpcResult, \rpcProof, \hashReq, \sigReq, \sigRes)
\]
where:
\begin{itemize}
    \item $\channelId$ is the identifier of the channel between $\LC$ and $\FN$.
    \item $\blockHeight$ is the current block height, indicating the time of response and also specifies which block header should be checked for $\rpcProof$. It has to be greater than or equal to the block height indicated in the request by $\req_i.\blockHash$.
    \item $\paymentAmount$ needs to match the input value $\req.\paymentAmount$ for $\hashReq$.
    \item $\rpcResult$ is the result for $\req.\rpcCall$, if applicable.
    \item $\rpcProof$ is the Merkle Proof of Inclusion of $\rpcResult$. 
    \item $\sigRes$ is the result of $\mathit{Sign}(\hashRes, \skfn)$, where \[\hashRes = \mathit{Hash}(\channelId, \blockHeight, \paymentAmount, \rpcResult, \rpcProof, \hashReq, \sigReq)\]
    
\end{itemize}


A unidirectional payment channel stored in the Channels Management Module (CMM) on the blockchain is denoted as $\PC$:
\[
\PC = (\channelId, \LC, \FN, \budget, \closingState, \status)
\]
where:
\begin{itemize}
    \item $\channelId$ is the unique identifier for the payment channel.
    \item $\LC$ is the address of the light client.
    \item $\FN$ is the address of the full node. 
    \item $\budget$ is the budget locked in this channel by $\LC$, limiting the maximum allowable $\req.\paymentAmount$.
    \item $\closingState$ represents the latest state of the payment channel, submitted by either participant and must be validated by a correct $\paymentAmount$ and $\sigAmount$ to be accepted.
    \item $\status$ denotes the status of the payment channel, with three possible values: $\mathit{Open}$, $\mathit{Closing}$, $\mathit{Closed}$.
\end{itemize}

In our scenario, $\LC$ only manages one payment channel $\PC$ locally, while $\FN$ and the CMM oversee several payment channels using identifiers with a mapping $(\channelId \mapsto \PC)$.


The channel state of a $\PC$ stored locally by $\LC$ and $\FN$ are the values of $\channelId$, $\paymentAmount$ and $\sigAmount$ exchanged in each round. Following the settlement of the channel, where the final channel state will be submitted on-chain, the funds are redistributed to the participants based on the payment amounts owed by $\LC$ to $\FN$. Moreover, $\res$ serves as a part of a fraud-proof sent by $\LC$ to the CMM to detect inconsistent returned data from $\FN$.


\subsection{State Transition of Participants}

The state transition diagram depicting the entire lifecycle of a PARP connection is illustrated in Figure \ref{fig:state}. 

A PARP-compatible full node can either be available or not available to a light client's payment channel connection request. To become available, a full node must deposit funds to the FNDM and indicate they are ready to serve. 

The state of a light client $\LC$ is deemed $\mathit{IDLE}$ if there is no established connection with a full node. It begins with the $\mathit{Handshaking}$ state to the $\mathit{Unbonded}$ state and subsequently to the $\mathit{Bonded}$ state upon establishing a payment channel. Later, it may enter the $\mathit{Unbonding}$ state if the connection is ending, ultimately returning to the $\mathit{IDLE}$ state. 

The state of a payment channel $\PC$ can be classified into three states: $\mathit{Open}$ means it is successfully set up; $\mathit{Closing}$ means one party wants to settle the channel, the channel will be under a period of time for disputes where the fund is time-locked; $\mathit{Closed}$ means it is successfully settled, and both parties receive the correct balance from the channel back.

\input{algorithms/handshake}

\begin{enumerate}
    \item \textbf{Initialization}: During this phase, $\LC$ seeks consent from $\FN$ to establish the payment channel $\PC$. In our design, where a full node is not required to deposit funds into a payment channel, mutual consent between $\LC$ and $\FN$ is crucial for channel creation. This confirmation includes an expiry time, indicating when the confirmation will expire, requiring $\LC$ to initiate another handshake if necessary. The algorithm is explained in Algorithm~\ref{alg:handshake}.
    \item \textbf{Channel Opening}: The creation of a payment channel $\PC$ is initiated by an $\mathit{OpenChannel}$ transaction sent from $\LC$, transitioning $\LC$'s state to $\mathit{Unbonded}$. This transaction includes metadata such as the budget amount $\PC.\budget$, participant addresses $\PC.\LC$ and $\PC.\FN$, and the signed confirmation from $\FN$. Additionally, it involves the transfer from $\LC$ of a certain amount of money equal to $\PC.\budget$. Upon receiving the $\mathit{TxReceipt}$ of the $\mathit{OpenChannel}$ transaction, $\LC$ transitions to the $\mathit{Bonded}$ state, and the channel's state managed by $\FN$ and CMM is set to be $\mathit{Open}$, with the identifier $\channelId$ assigned.
    
    \item \textbf{Active Phase}: This phase does not involve any on-chain participation. $\LC$ generates a $\req$ and sends it to $\FN$. $\FN$ then responds with a $\res$ and stores $\req.\paymentAmount$ and $\req.\sigAmount$. $LC$ stores the $\req.\paymentAmount$ locally and verifies $\res$.  
    \item \textbf{Termination}: Either party can send a $\mathit{CloseChannel}$ transaction which includes $(\alpha, \paymentAmount, \sigAmount)$ to the network. The action transitions $\LC$ to the $\mathit{Unbonding}$ state and $\PC$ to the $\mathit{Closing}$ state. 
    \begin{itemize}
        \item \textbf{No dispute.} If there is no dispute, which is the ideal case, after a period of dispute time, the channel is officially and successfully closed on-chain. CMM distributes $\PC.\budget$ accordingly based on $\paymentAmount$. The channel state managed by $\FN$ proceeds to $\mathit{Closed}$, while $\LC$ returns to $\mathit{IDLE}$. 
        \item \textbf{Dispute present.} Before the closure of the channel, either party can submit a final state different from the $\PC.\closingState$ recorded by CMM. The valid state with a higher value of $\paymentAmount$ will be acknowledged as the most recent state. Whenever a party submits a new valid latest state, the dispute time will be reset to allow the other party enough time to respond. The rest is the same as for the no-dispute scenario. 
    \end{itemize}

\end{enumerate}

\subsection{Liveness Check on the Payment Channel}

To facilitate a light client to monitor the payment channel's liveness, for example, if the payment channel is closed secretly by a full node, $\LC$ periodically sends a request to $\FN$ asking for $\PC.\status$. By getting block header information from other sources in the network as described in Section~\ref{subsec:assumption}, a light client can verify the liveness of a channel.


\subsection{Fraud Detection and Reporting}

To ensure system integrity, both the light client and the on-chain module perform a series of verification steps. The light client serves as the first line of defense, performing local checks to classify responses as \textbf{valid}, \textbf{invalid}, or \textbf{fraudulent}. If fraud is detected, it submits the relevant information as a fraud proof to the on-chain Fraud Detection Module, which then independently verifies the proof and enforces penalties. 

The light client verifies the response from the full node through the following checks:

\begin{itemize}
    \item \textbf{\textit{Verify Request Hash}}:
    To cryptographically link a PARP request with its response (which is needed to establish a fraud proof), the response must include the hash of its associated request (\emph{request hash}). The client must verify that the request hash matches the expected one. If not, the response is classified as \textbf{invalid} because the light client would not be able to generate a fraud proof, thus it cannot trust the response. If the hash matches but the request's signature fails verification, the response is also invalid. 
    \item \textbf{\textit{Verify Response Signature}}: The response message must contain a valid signature from the full node. If the signature is invalid, then again the client would not be able to generate a fraud proof to hold the full node accountable, so the response is classified as \textbf{invalid}.
    \item \textbf{\textit{Channel Identifier Check}}: The channel ID in the response must match the one in the request. Any mismatch is classified as \textbf{invalid}.
    \item \textbf{\textit{Payment Amount Check}}: The payment amount in the response must match the cumulative amount signed by the light client in the request. Any mismatch is \textbf{fraud}.
    \item \textbf{\textit{Timestamp Check}}: The block height in the response must not be lower than the one indicated in the request by the block hash. Any mismatch is considered \textbf{fraud}. 
    \item \textbf{\textit{Verify Merkle Proof}}: The response must contain a Merkle proof showing the data is part of the tree specified by the request type (e.g., transaction trie or state trie) at the current block height in the response. If the proof fails to verify, the response is considered \textbf{fraud}.
\end{itemize}

When the light client detects fraud, it submits the request $\req$, the response $\res$, and the address of the witness node $addr_\WN$ to the on-chain module. 


The on-chain module requires a trusted root hash $h_\mathit{root}$ for the relevant Merkle tree, determined by the return block height $\res.\blockHeight$. It will conduct the following checks:

\textbf{The integrity of the request:} It verifies $\req.\sigReq$ using $addr_\LC$ associated with $\channelId$ by reconstructing the hash value from the request contents.

\textbf{The origin of the response:} It verifies $\res.\sigRes$ using $addr_\FN$ associated with $\channelId$ by reconstructing the hash value from the response contents.

\textbf{The match of the identifier:} It verifies the identifier $\channelId$ in $\req$ matches the one in $\res$.

\textbf{The incorrectness of the response:} To penalize the full node, the module must confirm that $\res$ contains incorrect information. This includes any of the following scenarios:

\begin{itemize}
    \item Mismatch between the payment amount $\paymentAmount$ in $\req$ and $\res$.
    \item Outdated information due to return block height $\res.\blockHeight$ smaller than the one indicated by $req.\blockHash$.
    \item Invalid $\res.\rpcProof$ when verified against $h_\mathit{root}$.

\end{itemize}

\input{algorithms/fraud-proof}

The verification logic of a fraud-proof in the Channels Management Module (CMM) is shown in Algorithm~\ref{alg:fraudproof}.

%% file: algorithms/handshake.tex
\newcommand{\myin}{\Statex \hspace{1.5em}}
\algdef{SE}[Upon]{UponBlock}{EndUpon}[1]{\textbf{upon } #1}{}
\algtext*{EndUpon} 

\begin{algorithm*}
\caption{Light Client State Transition Logic during Initialization and Channel Opening Stages}
\label{alg:handshake}
\begin{algorithmic}[1]
\State \textbf{Initialization:}
    \myin Generate ($\pklc$, $\sklc$) and its address denoted as $\LC$
    \myin DECLARE $\blockHash$: a block HASH
    \myin DECLARE $\mathit{FN}$: an ADDRESS that represents a full node $\FN$
    \myin DECLARE $\mathit{step}$: one of the values in (IDLE, Handshaking, Unbonded, Bonded, Unbonding)
    \myin DECLARE $\alpha$: an INTEGER to identify a payment channel
    \myin DECLARE $\paymentAmount$: an INTEGER to record the usage of the budget, representing the latest local state of a channel

\UponBlock{start \textbf{call} StartHandShaking()}
\EndUpon

\Function{StartHandShaking}{}
    \State{$\blockHash$ $\gets$ Fetch the latest block hash from the network}
    \State{$FN \gets $\text{ Pick a full node}}
    \State{Send msg$\langle$HANDSHAKE, $\LC$$\rangle$ to $\FN$}
    \State{$step \gets$ Handshaking}
    
    \State{Set the $hsTimer$ timer}
\EndFunction

\UponBlock{receive msg$\langle$HSCONFIRM, $\pkfn$, $\mathit{expiryDate}$, $\mathit{Sign}((\LC||\mathit{expiryDate}), \skfn)$$\rangle$from $\FN$ while $\mathit{hsTimer}$ is active and $step = Handshaking$}
\State{Verify $\mathit{Sign}((\LC||\mathit{expiryDate}), \skfn)$ with $\pkfn$}

\State{\textbf{return} if signature is not valid}

\State{Form and Sign the Tx$\langle$OpenChannel($\mathit{Sign}((\LC||\mathit{expiryDate}), \skfn)$, $\pkfn$, $\pklc$, $\mathit{expiryDate}$$\rangle$}

\State{Attach the budget to Tx$\langle$OpenChannel$\rangle$}
\State{Send Tx$\langle$OpenChannel$\rangle$ to $\FN$}
\State{$step \gets$ Unbonded}

\EndUpon

\UponBlock{receive TxReciept$\langle$OpenChannel, $\mathit{Sign(channelId, \skfn)}, \mathit{channelId}$$\rangle$ from $\FN$ while $\mathit{step = Unbonded}$}
\State{Verify $\mathit{Sign(channelId, \skfn)}$ with $\pkfn$}
\State{$\alpha \gets channelId$}
\State{$\paymentAmount \gets 0$}
\State{$step \gets$ Bonded}
\EndUpon

\end{algorithmic}
\end{algorithm*}

%% file: algorithms/fraud-proof.tex
\newcommand{\reqDecoded}{\mathit{reqDec}}
\newcommand{\resDecoded}{\mathit{resDec}}
\newcommand{\channelInfo}{\mathit{chan}}
\newcommand{\reqHash}{\mathit{h_{req}}}
\newcommand{\resHash}{\mathit{h_{res}}}

\newcommand{\rootHash}{\mathit{h_{root}}}
\newcommand{\merkleHash}{\mathit{h_{merkle}}}

\newcommand{\require}{\textbf{require}}

\algdef{SE}[Upon]{UponBlock}{EndUpon}[1]{\textbf{upon } #1}{}
\algtext*{EndUpon} 

\begin{algorithm}
\caption{Fraud Proof Verification in the CMM}
\label{alg:fraudproof}
\begin{algorithmic}[1]

\Function{FraudProofDetection}{}
    \State \textbf{Input:} $\req, \res, addr_\WN$
    \State \textbf{Procedure:}


    $\reqDecoded$ $\gets$ decodeRequest(req)
    
    $\resDecoded$ $\gets$ decodeResponse(res)

    \textcolor{gray}{// The match of the identifier}

    $\require$($\reqDecoded.\channelId$ == $\resDecoded.\channelId$)

    $\channelInfo$ $\gets$ getChannelInfo($\reqDecoded.\channelId$)
    
    $\require$($\channelInfo.\status$ $\neq$ ``closed")

    \textcolor{gray}{// The origin of the request}

    $\reqHash$ $\gets$ hash($\reqDecoded$) 
    
    $\require$($\reqHash == \reqDecoded.\hashReq$)
    
    $\require$($\channelInfo.\LC$ == recover($\reqHash$, $\reqDecoded.\sigReq$))

    \textcolor{gray}{// The origin of the response}

    $\resHash$ $\gets$ hash($\resDecoded$) 
    
    
    $\require$($\channelInfo.\FN$ == recover($\resHash$, $\resDecoded.\sigRes$))




    \textcolor{gray}{// Payment Amount Check}
    \If{$\reqDecoded.\paymentAmount \neq \resDecoded.\paymentAmount$}  
        \State slashAndReward($\channelInfo.\FN$, $\channelInfo.\LC$, $addr_\WN$)
    \EndIf
    
    \textcolor{gray}{// Timestamp Check}\hspace*{\fill}

    $\mathit{m_{req}} \gets \text{getBlockHeightByHash}(\reqDecoded.\blockHash)$

    \If{$\resDecoded.\blockHeight < \mathit{m_{req}}$}
        \State slashAndReward($\channelInfo.\FN$, $\channelInfo.\LC$, $addr_\WN$)
    \EndIf

    \textcolor{gray}{// Verify Merkle Proof}
    \State $\rootHash \gets$ $\text{getRootHash}(\resDecoded.\blockHeight)$

    \If{$\text{verifyProof}(\rootHash, \resDecoded.\rpcProof) \neq true$}
        \State slashAndReward($\channelInfo.\FN$, $\channelInfo.\LC$, $addr_\WN$)
    \EndIf

\EndFunction

\end{algorithmic}
\end{algorithm}

%% file: Content/6-Implementation.tex
\section{Implementation and Evaluation}
\label{section6:implementation}


We developed a PARP prototype for Ethereum with three components: a full node, a light client, and on-chain modules. The full node, built on Geth (version 1.13.12 \cite{go-ethereum}), adds 1827 lines of Go (v1.20) for PARP compatibility. The light client uses around 1500 lines of Go. 1631 lines of Solidity (v0.8.25) manage deposits, payment channels, and fraud-proof detection. The prototype supports full interaction between components. We release our implementation for open science~\footnote{\url{https://github.com/podiumdesu/parp-dev}}.

The proof verification uses a Merkle proof. In our Ethereum-based implementation, Merkle Patricia Tries (MPT) generates a \textbf{state trie}, a \textbf{transaction trie}, and a \textbf{transaction receipt trie}, with root hashes stored in block headers. For the cost of submitting a fraud-proof, since Solidity cannot natively fetch root hashes for specific block numbers, the light client submits block header fields, including the trie roots, to regenerate the block hash on-chain. Ethereum's built-in block hash verification supports validation within the last 256 blocks, therefore, ensuring a trustworthy root hash.

We evaluated our prototype with these questions:

\begin{itemize}
    \item Additional communication costs of a PARP RPC request and its response in terms of message size? (Sec.~\textit{~\ref{section6-q1}})
    \item Additional processing time caused by PARP, and how does it impact request processing latency? (Sec.~\textit{~\ref{section6-q2}})
    \item On-chain costs incurred by PARP? (\textit{Sec.~\ref{section6-q3}})
    \item Additional CPU and memory are required by a PARP-compatible node compared to a standard Geth node as a function of the number of served clients? (\textit{Sec.~\ref{section6-q4}})
\end{itemize}

\subsection{Read and Write Workloads}

A \textbf{read} workload includes requests that query and retrieve data from the blockchain without altering its state. It is typical for data verification and status checks. 

A \textbf{write} workload refers to requests that change the state of the blockchain, typically by sending a signed transaction that will be validated and recorded on the blockchain. 

In PARP, each request and response includes base layer RPC data with additional PARP metadata for accountability. 

\subsection{Test Setup}

We established a local Ethereum network with three full nodes using Geth~\cite{go-ethereum} as the execution client. One node ran a PARP-compatible Geth version processing the connection and requests from a light client. The network was deployed on local OpenStack virtual machines with 4 vCPUs and 8 GB of RAM each, including a similarly configured light client.


\subsection{Communication Costs and Message Size}
\label{section6-q1}
\begin{figure}[htbp] 
    \centering 
    \includegraphics[width=0.9\linewidth]{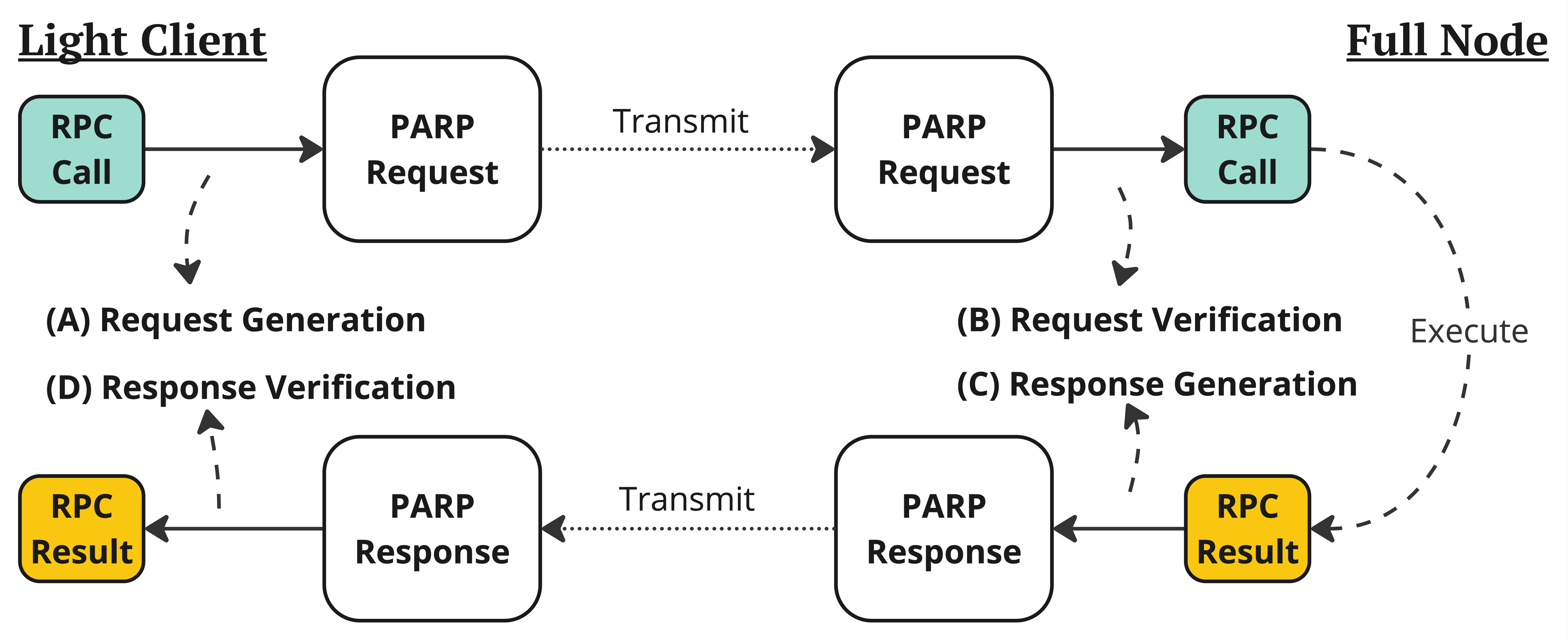}
    \caption{Additional computation steps required in processing a PARP request and response.}
    \label{fig:process-steps}
\end{figure}

As shown in Figure \ref{fig:process-steps}, an RPC call is wrapped with PARP metadata to form a PARP request sent to the full node. The full node processes it, executes the call, and wraps the result with response metadata before returning it to the light client.

For context, the size of an Ethereum JSON-RPC request for a raw transaction call, such as opening a payment channel, is \textbf{422} bytes, while retrieving an account balance is \textbf{118} bytes. 


A PARP request includes two 65-byte signatures for transaction integrity (returned in the response) and payment authentication. The total overhead per request is 226 bytes. 

A PARP response adds 187 bytes of metadata, including two signatures (one from the request), plus variable-sized proof verification data depending on the request type. The message size overheads are detailed in Table \ref{table:size_overhead}.

\input{table/benchmarking/size-overhead}

\begin{figure}[htbp] 
    \centering 
    \includegraphics[width=0.8\linewidth]{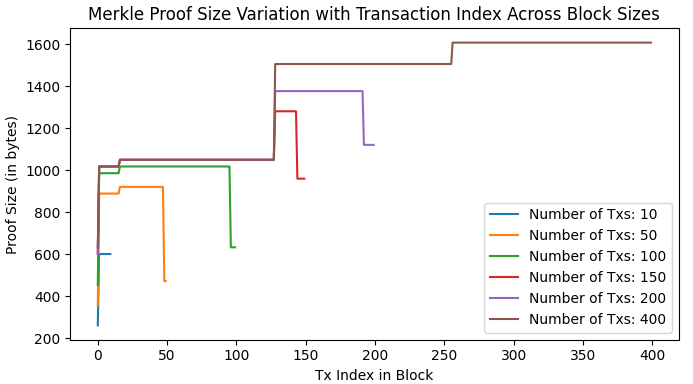}
    \caption{Merkle proof size variation with transaction index across a range of different block sizes. }
    \label{fig:mpt-size}
\end{figure}

We evaluated the size of Merkle proofs for transaction inclusion within blocks. Write requests use the transaction trie root. As shown in Figure \ref{fig:mpt-size}, Merkle proof sizes vary not only with \textbf{the number of transactions} included in one block but also with the \textbf{transaction index} within those blocks (explaining the sudden drop in the figure). For instance, for a transaction located in a block containing 200 transactions, the average Merkle proof size is approximately 1150 bytes.

\subsection{Computation Overhead and Latency}
\label{section6-q2}

Our protocol introduces additional steps and computational overhead compared to standard Geth interactions (Figure~\ref{fig:process-steps}). Table~\ref{table:time_costs} details the additional processing time of the steps marked (A) through (D). The reported numbers are the average increase in latency for 100 requests for both a write and a read workload. For the write workload, we generated a transaction in a block with 200 transactions. For the read workload, we use an RPC request that retrieves an account balance.

\input{table/benchmarking/latency.tex}

\input{table/benchmarking/on-chain-cost}
\subsection{On-chain costs}
\label{section6-q3}
\label{section6:on-chain-costs}

The PARP protocol includes several on-chain actions that incur costs in terms of gas fees. Table~\ref{table:on_chain_cost} summarizes the gas costs and their corresponding USD values on the Ethereum Mainnet and Arbitrum L2 network. At the time of calculation, we assumed an ETH price of \$4000 USD and gas prices of 12 Gwei for Ethereum Mainnet and 0.1 Gwei for Arbitrum.

\begin{figure}[htbp] 
    \centering 
    \includegraphics[width=\linewidth]{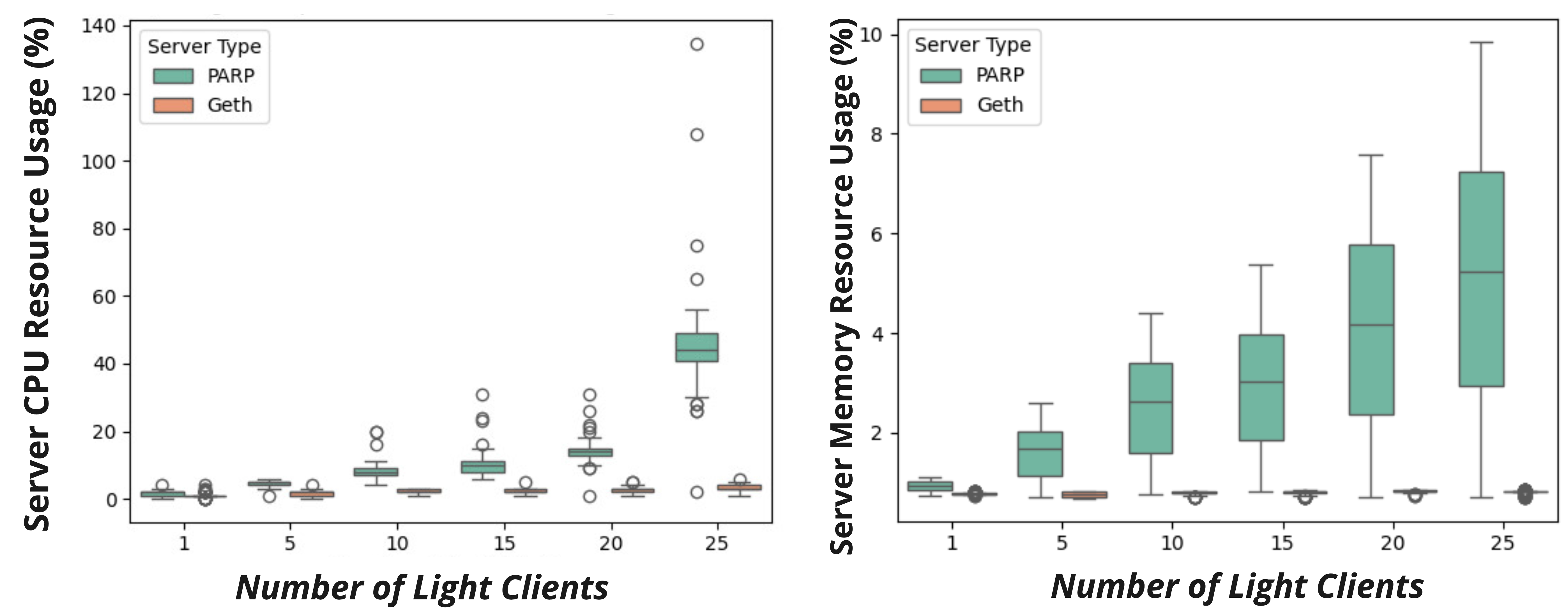}
    \caption{Additional computation steps required in processing a PARP request and response.}
    \label{fig:cpu-usage}
\end{figure}

\subsection{Scalability and Performance Metrics}
\label{section6-q4}
To evaluate scalability and performance, we tested a PARP-compatible full node with light clients sending two requests per second over two minutes. Using 4 vCPUs and 8GB RAM, the results in Figure~\ref{fig:cpu-usage}, indicate that when the number of simultaneous light client connections reached 20, the average CPU usage was 14.3\%, and the average memory usage was 2.63\%. Even though these numbers represent a 3.43x increase in CPU usage and a 2.38x increase in memory usage compared to a standard Geth node, we argue that they remain within an acceptable performance range and support our claim that PARP can be integrated into real-world blockchain networks with reasonable overhead. We expect future optimizations could further reduce resource consumption.


\subsection{Conclusion}
Our evaluation shows that PARP introduces manageable overhead while ensuring accountability and reliability in the blockchain RPC layer. Communication costs remain reasonable, and processing time adds only minor latency for both read and write operations. On-chain costs are transparent and can be reduced with Layer-2 solutions like Arbitrum. PARP-compatible full nodes maintain acceptable CPU and memory usage, even with multiple light clients. Overall, PARP integrates efficiently into blockchain networks without significant performance loss, enhancing incentives and trust.

%% file: table/benchmarking/size-overhead.tex
\begin{table}[H]
\centering
\begin{tabular}{|l|l|}


\hline
                         & \textbf{Size Overhead (in bytes)} \\ \hline
\textbf{PARP request}    & 226 bytes \\ \hline

\textbf{PARP response}   & 187 bytes + Size of Merkle Proof \\ \hline

\end{tabular}
\caption{Message Size overhead for PARP RPC requests and responses compared to standard Ethereum RPC calls.}
\label{table:size_overhead}
\end{table}

%% file: table/benchmarking/latency.tex
\begin{table}[H]
\centering
\begin{tabular}{|l|l|l|}
\hline
\textbf{Light Client Process Steps} & \textbf{Write} & \textbf{Read} \\ \hline
(A) Request Generation                      & 10.91ms & 4.82ms \\ \hline
(D) Response Verification (proof)    & 7.13ms & 5.78ms \\ \hline
(D) Response Verification (in total)        & 8.109ms & 1.01ms \\ \hline
\textbf{Full Node Process Steps} & \textbf{Write} & \textbf{Read} \\ \hline
(B) Request Verification                    & 714.43$\mu$s & 703.13$\mu$s  \\ \hline
(C) Response Generation (proof)      & 3.08ms & 477.12$\mu$s \\ \hline
(C) Response Generation (in total)          & 3.37ms & 1.29ms \\ \hline
\end{tabular}
\caption{Additional computational latency introduced by the protocol (average over 100 PARP RPC requests).}
\label{table:time_costs}
\end{table}

%% file: table/benchmarking/on-chain-cost.tex
\begin{table}[H]
\vspace{5pt}
\centering
\begin{tabular}{|m{1in}|c|c|c|}
\hline
\textbf{Action} & \textbf{Gas Cost} & \multicolumn{1}{|p{1cm}|}{\centering \textbf{MainNet} \\ \textbf{(USD)}} & \multicolumn{1}{|p{1.3cm}|}{\centering \textbf{Arbitrum} \\ \textbf{(USD)}} \\ \hline

\multicolumn{4}{|l|}{\textbf{Full Node Deposit}} \\ \hline
Deposit funds & 45238 & 2.171 & 0.018 \\ \hline

\multicolumn{4}{|l|}{\textbf{Channel Management}} \\ \hline

Open a channel & 196183 & 9.417 & 0.078 \\ \hline
Close a channel & 110118 & 5.286 & 0.044 \\ \hline
Confirm closure & 87128 & 4.182 & 0.035 \\ \hline

\multicolumn{4}{|l|}{\textbf{Fraud Proof Detection}} \\ \hline
Submit a fraud proof & 762508  & 36.6 & 0.305 \\ \hline

\multicolumn{2}{|l|}{\textit{\textbf{Median Transaction Fee (9/12/2024)}}} & \textit{1.606} & \textit{0.350} \\ \hline

\end{tabular}
\caption{On-Chain Cost Analysis of PARP Protocol. }
\vspace{-5pt}
\label{table:on_chain_cost}
\end{table}

%% file: Content/7-Related-work.tex
\section{Related Work}


Several projects share PARP's goal of making the blockchain RPC layer more decentralized and permissionless. The Portal Network ~\cite{portal} employs the Kademlia DHT~\cite{Maymounkov2002KademliaAP} to reduce the storage requirement for each node, allowing those with limited storage to contribute. This setup enables light clients to access specific network information on a peer-to-peer basis. However, the participation of nodes remains voluntary, with no mechanisms in place to reward serving nodes (in contrast to PARP's micro-payments approach). POKT~\cite{pokt} and Ankr~\cite{ankr} enable a network of decentralized nodes that serve requests in exchange for tokens but required to comply with Know Your Customer (KYC) regulations and to stake tokens. Ankr centralizes requests through a load balancer, while POKT requires users to access blockchain data through permissioned gateways. By contrast, PARP does not require the relay of requests through a centralized element and utilizes different incentives for node participation. 


RPCh~\cite{rpch} introduces a decentralized gateway where permissioned entry and exit nodes relay calls to the HOPR MixNet~\cite{hopr}. It incorporates ``Proof of Relay'' to reward nodes for effective data relay and uses payment channels for cost-efficient rewards. While RPCh offers enhanced privacy via a MixNet, PARP avoids routing calls through a permissioned entry node and uses distinct cryptoeconomic incentives.


Olshansky \emph{et. al}~\cite{olshansky2024relaymining} have proposed Relay Mining, where RPC nodes charge fees based on request volume, with rewards drawn from tokens staked by dApps. Selected RPC nodes serve a particular dApp for a designated period, spanning several blocks, after which they submit their proof of work on-chain through a commit-and-reveal scheme verified by Sparse Merkle Trees. In contrast, PARP uses cumulative payment amounts in payment channels to calculate work performed, establishing a different incentive model.

Among all these initiatives, PARP stands out as the only RPC protocol that supports verifiable blockchain data access using a fraud-proof protocol, with micro-payments as direct compensation for serving requests.

%% file: Content/8-Future-work.tex
\section{Limitations and Future Work}

\textbf{Risks of single node dependence.} In practice, a light client should connect multiple full nodes to enhance availability and avoid single points of failure. However, our protocol requires a light client to set up a payment channel individually with every full node it intends to connect with, adding costs and potentially discouraging multiple connections. Payment channel networks~\cite{nikolaos2020pcn} could address this by avoiding opening a dedicated channel per client-server pair.

\textbf{Privacy concerns in full node interactions.} Another limitation of our protocol is its inability to fully address concerns related to full nodes snooping on sensitive request information, including content and network information such as IP addresses. Future extensions may employ cryptographic methods like homomorphic encryption~\cite{smartfhe} and commitments~\cite{zether} for content privacy and Mixnets such as the Nym infrastructure~\cite{nym} for anonymous communication.

\textbf{Network rewards via Proof of Serving.} PARP can form a new reward mechanism that we tentatively call ``Proof of Serving', similar to Proof of Stake, but to incentivize full nodes to serve light clients. Payment proofs signed by light clients act as receipts, which full nodes can aggregate and submit to the network and claim a portion of the block reward. The main open issue is to address Sybil attacks whereby a full node controls fake light clients and connections. Introducing a reputation system to validate the legitimacy of served light clients could be one solution to this issue.

\textbf{Formalization of cryptoeconomic incentives.} We have not yet addressed the details of the economic incentives enabled by the PARP protocol.

Two key areas for PARP's economic incentives include designing a fee schedule for RPC requests and linking full node stakes to the volume of client requests they can handle. Fee schedules must balance client affordability with fair node compensation, while staking thresholds enhance security against misbehavior. Formalizing economic incentives to enhance honest participation in blockchain systems has been the subject of many studies~\cite{baldimtsi2023cryptoeconomic, deb2024stakesure, mamageishvili2023incentive}. Most relevant to our work Moshrefi \emph{et. al}~\cite{moshrefi2024unconditionally} introduce fraud-proof mechanisms with slashing conditions that penalize data tampering. The cost model employed in ``insured'' cryptoeconomic security varies with the value of the transaction, aligning financial risk with transaction importance. The incentive formation principles detailed in this paper can enhance our protocol by integrating a formalized compatible incentive model.



%% file: Content/9-Conclusion.tex
\section{Conclusion}

We have addressed the problem of the increasingly permissioned access to the serving layer of otherwise permissionless blockchain networks and its effects on the privacy, integrity, and availability of data access by application clients and end-users.
Our \textbf{P}ermissionless \textbf{A}ccountable \textbf{R}PC \textbf{P}rotocol (PARP) extends the RPC protocol of blockchain full nodes to enable pseudonymous yet accountable interaction between light clients and full nodes. The protocol provides an alternative to permissioned and privacy-invasive but reputable NaaS providers on the one hand, and permissionless but less reliable public RPC nodes on the other hand. Our implementation and evaluation of the PARP protocol for the Ethereum network demonstrates the feasibility of a light client obtaining RPC service from a full node with high data security for the client, ensured payment for the full node, and acceptable computation and communication overhead for both parties.

